\documentclass{JHEP3}

\usepackage{epsfig}
\usepackage{amssymb}

\def\al{\alpha}

\def\om{\omega}

\def\map{\rightarrow}
\def\bq{\begin{equation}}
\def\eq{\end{equation}}
\def\li{\langle}
\def\ri{\rangle}

\def\ti{\tilde}
\def\Om{\Omega}
\def\Si{\Sigma}
\def\La{\Lambda}
\def\ss{\subset}
\def\ovl{\overline}
\def\Hi{{\cal H}}
\newtheorem{theorem}{Theorem}
\newtheorem{cor}{Corollary}
\newtheorem{conjecture}{Conjecture}

\title{(Non-)Abelian Kramers--Wannier Duality\\ And Topological
Field Theory}

\author{Pavol \v Severa \thanks{work supported by EPDI} \\ Dept.~of Theoretical Physics\\Comenius University
\\Mlynsk\'a dolina \\84215 Bratislava, \\Slovakia \\\email{severa@sophia.dpt.fmph.uniba.sk}
}

\abstract{
We study a connection between duality and topological field theories. First, 2d Kramers--Wannier
duality is formulated as a simple 3d topological claim (more or less Poincar\'e duality), and a similar
formulation is given for higher-dimensional cases. In this form they lead to
simple TFTs with boundary coloured in two colours. The statistical
models live on the boundary of these TFTs, as in the CS/WZW or AdS/CFT correspondence.

 Classical models (Poisson--Lie T-duality) suggest a non-abelian generalization in the 2d
case, with abelian groups replaced by quantum groups. Amazingly, the TFT
formulation solves the problem without computation: quantum
groups appear in pictures, independently of the classical motivation. Connection with Chern--Simons theory appears at the symplectic level, and also in the pictures of
the Drinfeld double: Reshetikhin--Turaev invariants of links in 3-manifolds, computed from the double, are included in these TFTs. All this suggests nice phenomena in higher dimensions.}

\begin{document}

\section{Introduction: KW duality as a 3-dimensional topological claim}
Kramers--Wannier duality in 2d statistical models is a rather elementary
equivalence between two models defined on mutually dual planar graphs. Yet it
may be a bit surprising that one can give a two-line proof of its most general
form (including all topological subtleties, order-disorder operator duality, etc.),
together with its higher-dimensional
generalizations. The basic idea is, roughly speaking, to consider surfaces that
are boundaries of 3d bodies (or $n$-manifolds, boundaries of $n+1$-manifolds).
As we shall see, the 2d models ``live on the boundary'' of a 3d topological
field theory---an idea familiar from WZW/Chern--Simons correspondence, or from
AdS/CFT duality. In this paper we try to study non-Abelian generalizations of
KW-duality, following this connection with TFT.

The duality is a claim about pictures like this:
\EPSFIGURE{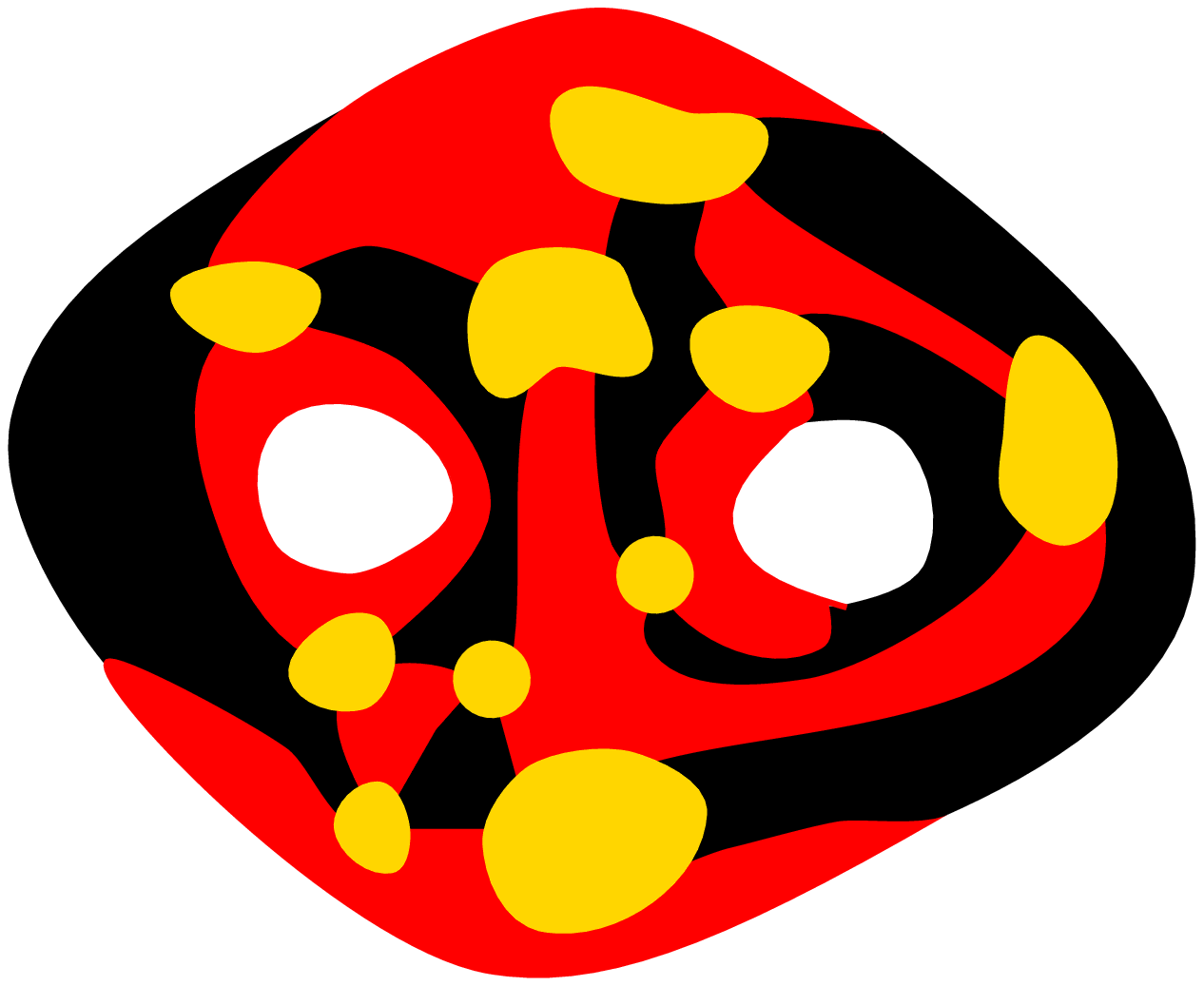, width=4cm}{A mask}
\noindent The picture represents a 3d body (a ritual mask) made of yellow material,
with the surface partially painted in  red and black. For definiteness
imagine that the invisible side is unpainted (i.e. completely yellow).
In general we have a compact oriented yellow $n+1$-fold $\Om$ with the
boundary coloured in this way (in a locally nice way; more precisely, $\Om$ is
a cobordism connecting yellow surfaces with boundaries colored in red and black).

We decompose $n$ as $n=k+l$ ($2=1+1$ in 2d statistical models)
and choose a finite abelian group $G$ and its dual $\ti G$. Let $y$ be the
yellow
 part of the boundary; it is a compact oriented $n$-dim surface with the boundary
coloured in black and red. The relative cohomology groups $H^k(y,r_y;G)$ and
$H^l(y,b_y;\ti G)$ (both are finite Abelian groups) are mutually dual via
Poincar\'e duality ($r_X$ denotes the red part of $X$; notice that $r_y=r_\Om\cap y$). Let
$\rho:H^k(\Om,r_\Om;G)\map H^k(y,r_y;G)$ and $\ti\rho:H^l(\Om,b_\Om;\ti G)\map
H^l(y,b_y;\ti G)$ be the restriction maps.

\begin{theorem}(Kramers-Wannier duality)\\
The images of $\rho$ and $\ti\rho$ are each other's annihilators.
\end{theorem}
\Proof{It is an immediate consequence of a piece of the long exact sequence of
the triple $r_\Om\ss y\cup r_\Om\ss\Om$,
$$H^k(\Om,r_\Om;G)\map H^k(y\cup r_\Om,r_\Om;G)\map H^{k+1}(\Om,y\cup r_\Om;G),$$
where the first arrow is (by excision) $\rho$ and the second (by Poincar\'e duality)
is the adjoint of $\ti\rho$, q.e.d.}

In statistical models it is used in the following form: for any function
$f:H^k(y,r_y;G)\map\mathbb{C}$ (the Boltzmann weight) we define the partition sum
\bq Z(f)=\sum_{x\in H^k(\Om,r_\Om;G)}f(\rho(x)).\eq
Let $\hat f$ denote the Fourier
transform of $f$. We can define \bq\ti Z(\hat f)=\sum_{x\in H^l(\Om,b_\Om;\ti
G)}\hat f(\ti\rho(x)).\eq   Theorem 1 and Poisson summation
formula now give
\begin{cor}(KW duality, usual form)\\For some constant $c$ (independent of $f$), $Z(f)=c\ti Z(\hat f)$.
\end{cor}

Let us stop to make a connection with more usual formulations. Notice that an element
of $H^1(X,Y;G)$ is the same as (the isomorphism class of) a principal $G$-bundle
over $X$ with a given section over $Y\ss X$. If $\Om$ is a 3d ball (with
coloured surface), an element of $H^1(\Om,r_\Om;G)$ is therefore specified by
choosing an element of $G$ for each red stain. We may imagine that there is a
$G$-valued spin sitting at each such  stain and, to compute (1.1), we take the sum
over all their values (we overcount $|G|$ times, but it is inessential). According to KW
duality, the same result can be obtained by summing over $\ti
G$-spins at the black stains. The spins at red (or black) stains interact through the yellow stains. If all the yellow stains are as those visible on the picture (disks with two red and two black neighbours), we have the usual two-point interactions; for disks with more neighbours we would have more-point interactions.

Finally, let us look at the picture again. It does not represent a ball and
the
back yellow stain is not a disk. The Boltzmann weight for the back stain can
be
understood as the specification of the boundary and periodicity conditions on
the visible surface (the $G$-bundle type together with sections over the red
 parts of the boundary). The boundary and periodicity conditions at the holes
 can be interpreted as order and disorded operators respectively.

These examples are more or less all that we would like; the general case seems
to be general beyond usual applications. But it will come handy when we consider non-abelian generalizations.

What are we going to do? First of all,  expression (1.1) has the form of a very
simple topological field theory (with boundary coloured in red and black), described in the next section. We shall then
look at the non-abelian version. In the $2=1+1$ case classical models suggest
that the pair $G$, $\ti G$ should be replaced by a pair of mutually dual quantum
groups. So we are faced with the difficult and somewhat arbitrary task of defining
and understanding quantum analogues of cohomology groups and of the Poisson summation
formula. But miraculously, none of these has to be done. Pictures alone (in the form of TFTs)  solve
the problem and quantum groups appear. This suggests, of course, that this point
of view might be interesting in higher dimensions, the $4=2+2$ case---the
electric--magnetic duality---being of particular interest.

\section{KW TFTs and the squeezing property}
As we mentioned,  expression (1.1) has the
form of a TFT with boundary coloured in red and black.
We understand TFT as defined by Atiyah \cite{Ati} and for definiteness
we choose its hermitian version; nothing like central extensions is
taken into account. To each oriented yellow $n$-dim $\Si$ with black-and-red
boundary, we associate a non-zero finite-dimensional Hilbert space $\Hi(\Si)=L^2(H^k(\Si,r_\Si;G))$. And for
each $\Om$ we have a linear form on the Hilbert space corresponding to $y$  --
the one given by (1.1). However, the normalization has to be changed slightly for
the gluing property to hold (this is only a technical problem): we set
\bq Z_\Om(f)={1\over\mu(\Om)}\sum_{x\in H^k(\Om,r_\Om;G)}f(\rho(x))\eq
and for the inner product
\bq \li f,g\ri=\mu(\Si)\sum_{x\in H^k(\Si,r_\Si;G)}\ovl{f(x)}g(x).\eq
Here
\bq \mu(\Om)={|H^{k-1}(\Om,r_\Om;G)||H^{k-3}(\Om,r_\Om;G)|\dots\over
|H^{k-2}(\Om,r_\Om;G)||H^{k-4}(\Om,r_\Om;G)|\dots}\eq
(and the the same for $\Si$). Perhaps this $\mu$ is not a number you would like
to meet in a dark forest, but this should not hide the simplicity of the thing.
The gluing property follows from the exact sequence for the triple
$r_{glued}\ss\Om\cup r_{glued}\ss\Om_{glued}$ ($r_{glued}$ is the red part of
$\Om_{glued}$; $\Om\ss\Om_{glued}$ is achieved by separating slightly the glued
yellow surfaces). Of course, the expression for $\mu$ was actually derived from
this sequence. Hence we can state
\begin{theorem} The asignment $\Si\mapsto\Hi(\Si)$ and $\Om\mapsto Z_\Om$ is
a TFT (for cobordisms with boundary colored in two colors).\end{theorem}

This TFT reformulation of KW duality will be our starting point
for non-abelian generalizations (the duality is an isomorphism between the TFT
given by $G$ and $k$, and the TFT given by $\ti G$ and $l$, with exchanged
red and black). Let us first have a look to see if we can
recover the numbers $k$ and $l$ and the group $G$ from the TFT. It is
enough to take yellow $n$-dim balls
as $\Si$'s. The ball should be painted as follows:
let us choose integers $k',l'$ such that $k'+l'=n$; we take a
$S^{k'-1}\ss\partial\Si$ and paint its tubular neighbourhood in $\partial\Si$ in
red; the rest (a tubular neighbourhood of a $S^{l'-1}$) is in black. Let us
denote this $\Si$ as $B^n_{k',l'}$. The corresponding Hilbert space $\Hi(B^n_{k',l'})$
is trivial
(equal to $\Bbb C$) if $k'\ne k$; if $k'=k$, it is the space of functions on $G$.
The reader may try to define the Hopf algebra structure on this space using
pictures (the $2=1+1$-case is drawn in the next section).

Our TFTs are of a rather special nature, because of the excision property of relative cohomology. It gives rise to the {\it squeezing property} of our TFTs. It is
best explained by using an example. Imagine this full cylinder (the upper half of its mantle is red and the lower half is black; the invisible base is yellow):

\EPSFIGURE{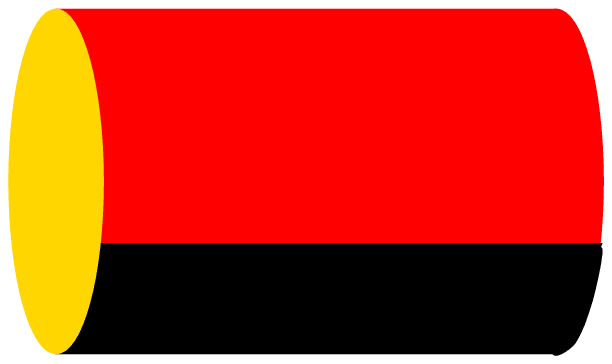, width=4cm}{before...}
We shall squeeze it in the middle, putting one finger on the red top and the
other on the black bottom. The result is no longer a manifold---it has a
rectangle in the middle (red from the top and black from the bottom), but it is
surely homotopically equivalent (as a pair $(\Om,r_\Om)$, or as a pair $(\Om,b_\Om)$) to the original cylinder. Since we use relative cohomologies, the
rectangle may be removed (it does not matter whether the cohomologies are
relative to $r$ or to $b$ (the dual picture), as the rectangle is
both red and black). The result is again a manifold of the type we admit:

\EPSFIGURE{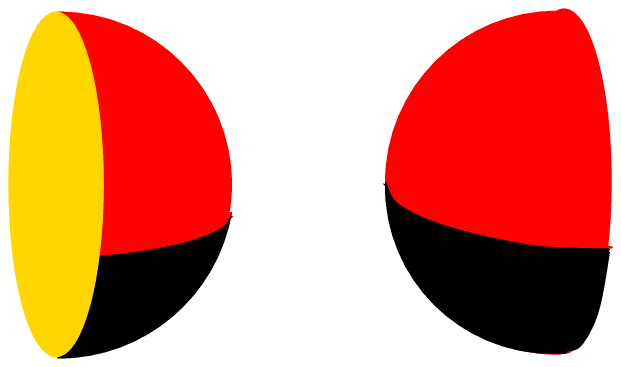, width=4cm}{...and after}

Or, as another example: if our fingers are not big enough, we do not separate the cylinder into two
parts, but instead we produce a hole in the middle (the top view of the result
would be a red stain with a hole in the middle).

A bit informally the squeezing
property can be formulated as follows: if a (hyper)surface appears as a result of squeezing $\Om$, red from one side and   black  from the other side, it
may be removed. Using excision, we can state
\begin{theorem}
The TFTs of Theorem 2 satisfy the squeezing property.
\end{theorem}

Those TFTs that satisfy the squeezing property may be considered as
generalizations of relative cohomology and of KW duality. As we shall
see in the next setion, in the $2=1+1$-case they yield the expected result.

{\bf Example:}
Here is an example of such a TFT that does not come from an abelian group.
We take
a finite group $G$ and  two subgroups $R,B\ss G$ such that
$RB=G$, $R\cap B=1$.  We shall consider principal $G$-bundles
with reduction to $R$ over $r$ and to $B$ over $b$. If $P$ is such a thing, let
$\mu(P)$ be the number of automorphisms of $P$. If $M$ is a space with some red
and some black parts, let $P(M)$ be the set of isomorphism classes of these
things. We
set $\Hi(\Si)$ (the Hilbert space) to be the space of functions on $P(\Si)$ with
the inner product \bq\li f,g\ri=\sum_{P\in P(\Si)}\mu(P)\overline{f(P)}g(P)\eq
and finally, if $f\in\Hi(y)$, we set
\bq Z_\Om(f)=\sum_{P\in P(\Om)}{1\over\mu(P)}f(P|_y).\eq
This is surely a TFT. The squeezing property holds, because if we have a
reduction for both $R$ and $B$ (as we have on the surfaces that appear by
squeezing), these two reductions intersect in a section of the $G$-bundle. If
$R=1$ and $B=G$, this TFT describes interacting $G$-spins (as in the
introduction); the general case is more interesting, providing a
non-trivial example for \S3.  We will also meet its version in \S4.

\section{Non-abelian $2=1+1$-duality}

There are classical models (those appearing in Poisson--Lie T-duality \cite{PL})
that suggest a non-abelian generalization of $2=1+1$ KW duality. The PL T-duality
generalizes the usual $R\leftrightarrow1/R$ T-duality, replacing the two circles
(or tori) by a pair of mutually dual PL groups. Clearly, we have to replace the
pair $G$, $\ti G$ by a pair of mutually dual quantum groups. This is not an easy
(or well-defined) task. We have to define and to {\it understand}
cohomologies with quantum coefficients.

\EPSFIGURE{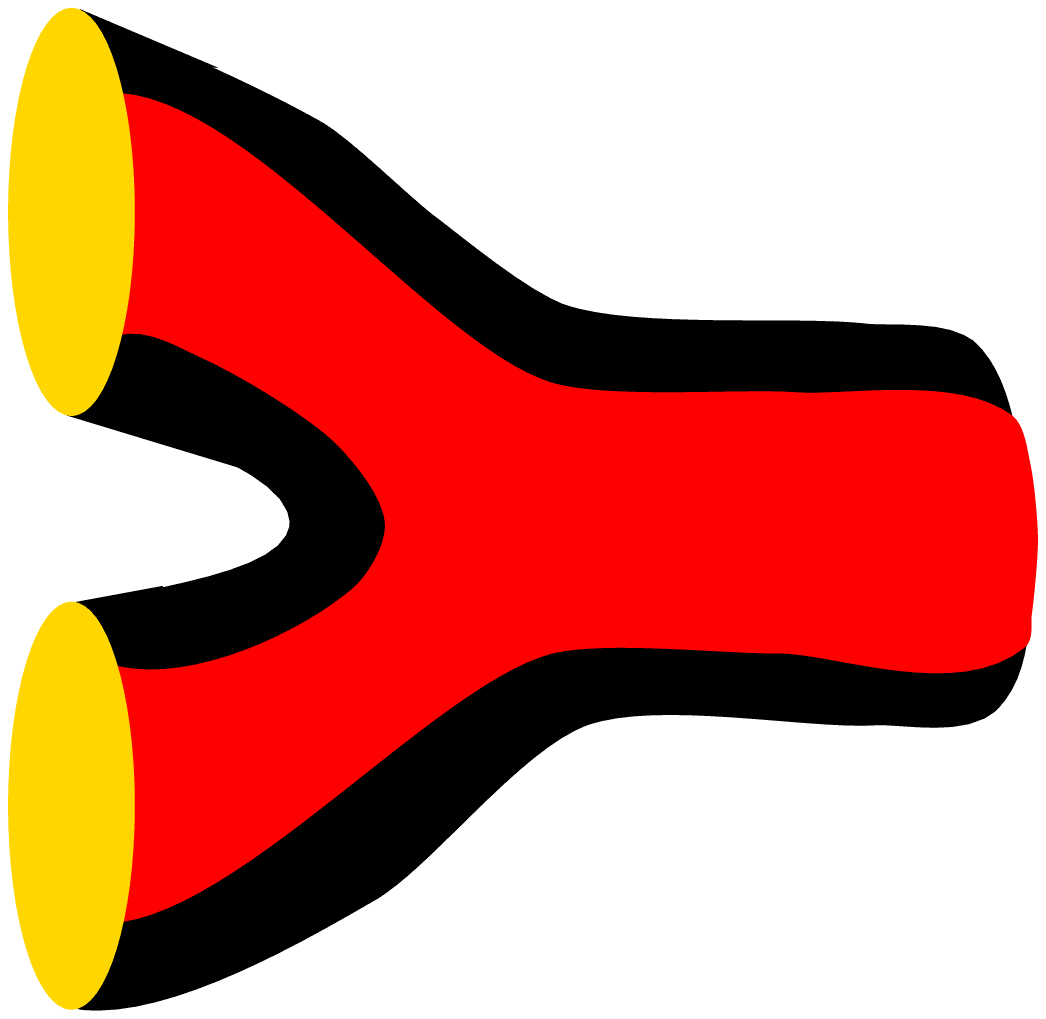, width=3.7cm}{Product}
Here is how pictures solve this problem in a very simple way: just take a TFT in three dimensions,
satisfying the squeezing property. A finite quantum group (finite-dimensional Hopf $C^*$-algebra) will appear
independently of the classical motivation. If you exchange red and black (which
gives a new TFT), the quantum group will be replaced by its dual. This is the
non-abelian (or quantum) $2=1+1$ KW duality.

Now we will draw the pictures. I learned this 3d way of representing quantum
groups at a lecture by Kontsevich \cite{Kon}; it was one of the sources of this
work. The finite quantum group itself is $\Hi(B^2_{1,1})$. The product
$\Hi(B^2_{1,1})\otimes\Hi(B^2_{1,1})\map\Hi(B^2_{1,1})$ is on fig. 4.

\EPSFIGURE{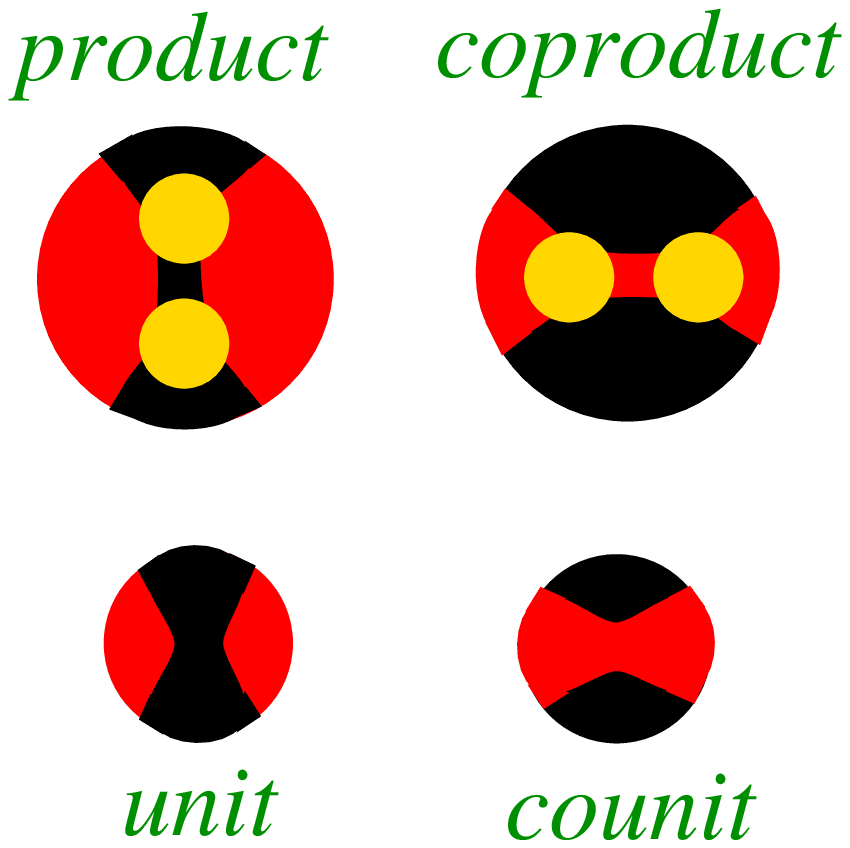, width=3.7cm}{All operations}
And here are all the operations. Coloured 3d objects are hard to draw (but not
hard to visualize!); imagine that the pictures represent balls and that their
invisible sides are completely yellow. The antipode $S$ is simply the half-turn, the
involution $*$ is the reflection with respect to the horizontal diameter, and
the rest is on the figure:

Why is it a quantum group? Just imagine the pictures representing the
finite quantum group axioms and
use the squeezing property in a very simple manner. We (more precisely, you) have
proved
\begin{theorem} A 3d TFT satisfying the squeezing property yields a finite
quantum group.
\end{theorem}

\begin{conjecture}There is a 1--1 correspondence between finite
quantum groups and 3d TFTs satisfying the squeezing property, with trivial
(i.e. one-dimensional) $\Hi(B^2_{0,2})$ and $\Hi(B^2_{2,0})$.
\end{conjecture}

To support
the conjecture, finite quantum groups are in 1--1 correspondence with modular
functors of a certain kind (cf. \cite{KW}), clearly connected with our TFTs.

\section{Chern--Simons with coloured boundary}

Let us recall a basic analogy between symplectic manifolds and vector spaces (the aim of quantization is to go beyond a mere analogy):

\begin{center}
\begin{tabular}{|c|c|} \hline
{\em Vector}&{\em Symplectic}\\ \hline\hline
Vector space&Symplectic manifold\\ \hline
Vector&Lagrangian submanifold\\ \hline
$V_1\otimes V_2$&$M_1\times M_2$\\ \hline &\\[-2.5ex]
$V^*$&$\overline{M}$\\ \hline
Composition of linear maps&Composition of Lagrangian relations\\ \hline
\end{tabular}
\end{center}

\def\fg{{\frak g}}
\def\fb{{\frak b}}
\def\fr{{\frak r}}
One can easily describe the symplectic analogue of the Chern--Simons TFT (see e.g. \cite{Fr}). Let $\fg$ be a Lie algebra with invariant inner product. If $\Si$ is a closed oriented surface then the moduli space of flat $\fg$-connections is a symplectic manifold (with singularities). The symplectic form is given as follows. The vector space
of all $\fg$-valued 1-forms on $\Si$ is symplectic, with the symplectic form
\bq\om(\al_1,\al_2)=\int_\Si\langle\al_1,\al_2\rangle.\eq
When we restrict ourselves to flat connections, the space is no longer symplectic, but the null directions of the 2-form give just the orbits of the gauge group, so the quotient (the moduli space) is symplectic. Let us denote it by $M_\Si$.

We have associated a symplectic space to every oriented closed surface.
Now, if $\Om$ is an oriented compact 3-fold with boundary $\Si$, we should
find a Lagrangian subspace $\La_\Om\ss M_\Si$. Indeed, $\La_\Om$ consists
just of those flat connections on $\Si$, which can be extended to $\Om$.

Let us make a minute extension of this construction, allowing a boundary
coloured in red and black. Let $\fb,\fr\ss\fg$ be a Manin triple. We shall
consider flat $\fg$ connections as before, with the obvious boundary
conditions---on the red part of the boundary the connection should take
values in $\fr$ and on the black part in $\fb$. Similarly, the gauge group
consists of the maps to $G$ with the same boundary conditions. This really
defines a symplectic TFT for our pictures. From this symplectic TFT we obtain
a symplectic analogue of quantum group (using the pictures of the previous
section). One readily checks that it is the double symplectic groupoid of
Lu and Weinstein \cite{LW}---the symplectic analogue of the quantum group
coming from the Manin triple $\fb,\fr\ss\fg$.

For this reason, it is reasonable to conjecture that perturbative
quantization of our Chern--Simons TFT with boundary will give the
corresponding quantum group.

In the next section we  return to the vector side of our table, to general
3d TFTs that satisfy the squeezing property. We shall see this connection
with CS TFTs again, in a different guise.

\section{Pictures of the Drinfeld double}

\EPSFIGURE{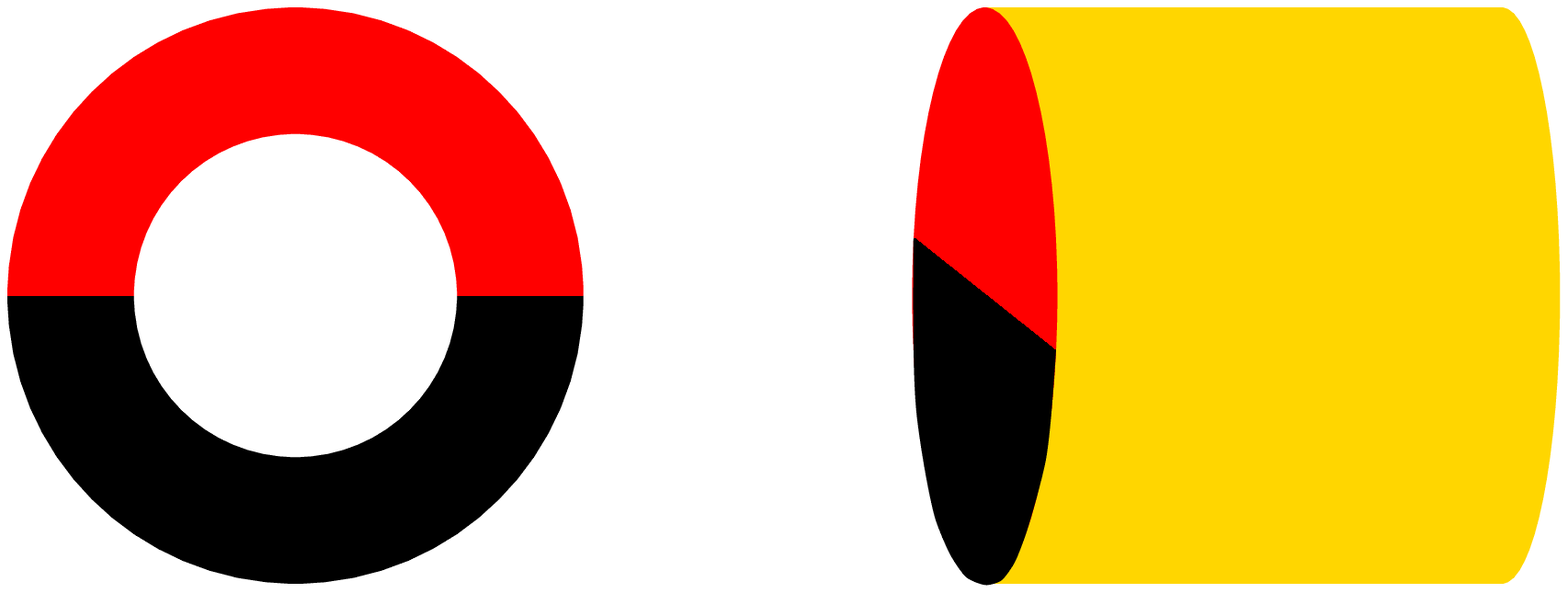, width=5.5cm}{Double unit \& counit,}
There are lots of algebras, modules,  etc., in our pictures. We shall describe
only the Drinfeld double, since it is important in PL T-duality, and also
to make a connection with Reshetikhin--Turaev invariants. Here are the unit and the
counit:

The invisible side of the full torus on the first picture is yellow; this closed
yellow strip is the double.
\EPSFIGURE{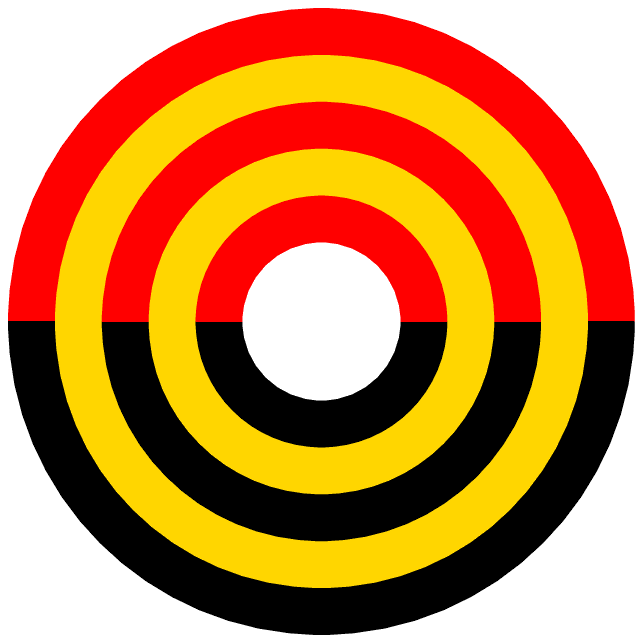, width=3cm}{product,}
\noindent On the second picture it is represented as the
mantle of the cylinder (the invisible base of the cylinder is painted as the visible one).

The product (the picture is yellow from the invisible side) and coproduct are on
figures 7 and 8.
The last picture requires an explanation. It represents a thick Y from which a thin
Y was removed (you can see it as the black holes in the yellow disks). The
fronts of these Y's are red and their backs are black (the invisible bottom of
the picture is yellow---it is the third double).

\EPSFIGURE{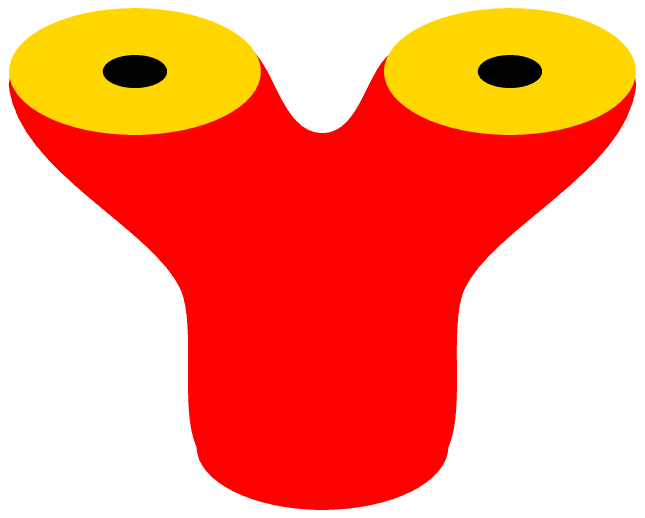, width=3cm}{\\ coproduct,}
For completeness, the antipode is a half-turn and the involution a reflection,
both exchanging the boundary circles of the double.

\font\cyr=wncyr10

Now we know the double as a Hopf algebra, but its real treasure is the
$R$-matrix.
It is quite similar to the Y-picture, but this time we do not remove a thin X,
but rather two tubes connecting the top holes with the bottom ones. However, if
one tube connected the left holes and the other one the right holes, the picture
would not be very interesting. We could squeeze the X in the middle, dividing it
into two vertical cylinders. We would simply have an identity.
\EPSFIGURE{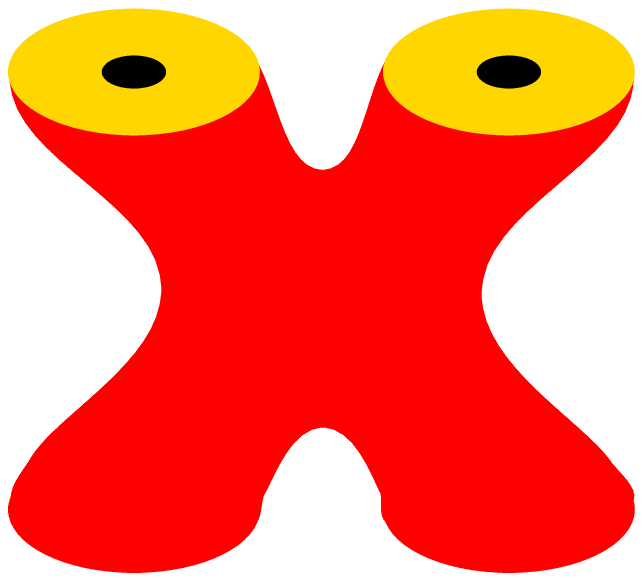, width=3cm}{and\\ $R$-matrix.}
However, in the X of the $R$-matrix, the tubes are diagonal. There are two ways for them to avoid each
other; one gives the $R$-matrix and the other its inverse.
This X has two incoming and two outgoing doubles; you can also imagine $n$
doubles at the bottom, tubes forming a braid inside and leaving the body at the
top, in the middle of $n$ other doubles (the Cyrillic letter {\cyr ZH} is good here). We  directly see a representation of the braid group.

With this picture in mind, we can find the Reshetikhin--Turaev (RT)
invariants coming from the double. Namely, we have

\begin{theorem} The boundary-free part of a 3d TFT satisfying the squeezing
property is the Chern--Simons TFT for the corresponding Drinfeld double.
\end{theorem}

Here is a sketch
of the proof: suppose $\Om$ is a closed oriented 3-fold with a ribbon link.
We colour each  of the ribbons in red on one side and in black on the other
side, blow it a little, so that the ribbon becomes a full torus removed from
$\Om$, and paint on the torus a little yellow belt. Our TFT gives us an
element of ${\rm double}^{\otimes n}$ (one double for each yellow belt),
where $n$ is the number of components of the link. Actually, this element is
from $(\mbox{centre of double})^{\otimes n}$ (we can move a yellow belt
along the torus and come back from the other side). It is equal to the RT
invariant. This claim follows immediately from the definition of RT
invariants: If $\Om=S^3$, we are back in our picture of braid group, and
generally, surgery along tori in $S^3$ can be replaced by gluing tori along
the yellow belts.

Finally, we can get rid of red and black and instead consider $\Om$'s with
boundary consisting of yellow tori: one easily sees that
$\Hi(\mbox{yellow torus})= \mbox{centre of double}$, q.e.d.

\section{Conclusion: Higher dimensions?}

There are several open problems remaining. Apart from the mentioned
conjectures there is a problem with the square of the antipode: for the
naive definition of TFT used in this paper, it has to be 1. One should find
a less naive definition and prove in some form the claim that our pictures
are equivalent to Hopf algebras.

However, in spite of these open problems, the presented picture is very simple and quite appealing. It is really tempting
(and almost surely incorrect) to suggest \bq\mbox{\it duality = TFT with the
squeezing property.}\eq It would be nice to understand the basic building blocks
of these TFTs that replace quantum groups in higher dimensions. It is a purely
topological problem. It would also be nice to have a non-trivial example with
non-trivial $\Hi(B^4_{2,2})$, to see an instance of S-duality ($4=2+2$ duality) in this way.

The field of duality is vast and connections with this work may be of diverse
nature. But let us finish with a rather internal question: Why yellow, red
and black?

\end{document}